\documentclass[a4paper,twocolumn,11pt,accepted=2025-03-28]{quantumarticle}
\pdfoutput=1
\usepackage[utf8]{inputenc}
\usepackage[english]{babel}
\usepackage[T1]{fontenc}
\usepackage{amsmath}
\usepackage{amssymb}
\usepackage{txfonts}
\usepackage{pifont}
\usepackage[colorlinks]{hyperref}
\usepackage{tikz}
\usepackage{lipsum}
\usepackage{graphicx}
\usepackage{verbatim}

\usepackage[numbers]{natbib}
\usepackage{ragged2e}
\usepackage{color}
\usepackage{float}
\usepackage{soul}

\usepackage{bm}
\usepackage{dcolumn}
\usepackage{xr-hyper}

\begin{document}

\title{Controlling measurement-induced phase transitions with tunable detector coupling}

\author{Ritu Nehra}
\affiliation{Department of Physics, Ben Gurion University of the Negev, Israel}
\email{ritu@post.bgu.ac.il}
\orcid{0000-0002-8983-2588}

\author{Alessandro Romito}
\affiliation{Department of Physics, Lancaster University, United Kingdom}
\email{alessandro.romito@lancaster.ac.uk}
\orcid{0000-0003-3082-6279}

\author{Dganit Meidan}
\affiliation{Department of Physics, Ben Gurion University of the Negev, Israel}
\affiliation{Université Paris-Saclay, CNRS, Laboratoire de Physique des Solides, 91405, Orsay, France.}
\email{dganit@bgu.ac.il}
\orcid{0000-0002-6263-5568}

\maketitle

\begin{abstract}

We study the evolution of a quantum many-body system driven by two competing measurements, which induces a topological entanglement transition between two distinct area law phases. We employ a positive operator-valued measurement with variable coupling between the system and detector within free fermion dynamics. This approach allows us to continuously track the universal properties of the transition between projective and continuous monitoring. Our findings suggest that the percolation universality of the transition in the projective limit is unstable when the system-detector coupling is reduced.

\end{abstract}

\section{ Introduction}
In recent years, there has been a significant surge of interest in investigating the spread of quantum information within open many-body quantum systems. This interest spans areas like quantum computation and teleportation,  condensed matter theory, and statistical physics.  
The environment acts as a measurement apparatus, continuously probing the system~\cite{jacobs_quantum_2014,Quantuminformation}. These repeated measurements restrict the entanglement growth and corrupt the quantum information,  leading to a measurement-induced phase transition marked by a sudden shift in the scaling of entanglement entropy~\cite{eisert_colloquium_2010,PhysRevX.7.031016,PhysRevB.98.205136,PhysRevB.99.224307,skinner_measurement-induced_2019,szyniszewski_entanglement_2019,Altmanerror2020,Bao2020,fisher_random_2023}, reported in cutting edge experiments\cite{noel2022measurement,google2023measurement,koh2023measurement}.

Entanglement transitions can also arise in a system subjected to competing measurements.  Here, the competition between the two measurements drives a topological entanglement transition between distinct disentangled phases.  This procedure can be used to tailor quantum states with distinct topological order~\cite{Lang2020,Roy2020,lavasani_measurement-induced_2021, lavasani_topological_2021,Ippoliti2021entanglement,Sang2021measurement,kells_topological_2023,klocke2024entanglementdynamicsmonitoredkitaev} which can be potentially useful for quantum error correction\cite{KITAEV2003,kitaev_anyons_2006}. The critical properties of topological entanglement transition under projective measurement differ considerably from those driven by continuous weak monitoring
~\cite{lavasani_measurement-induced_2021,PhysRevX.13.041028,kells_topological_2023,PhysRevX.13.041045}.
\begin{figure}[h!]
\centering
\includegraphics[width=0.475\textwidth]{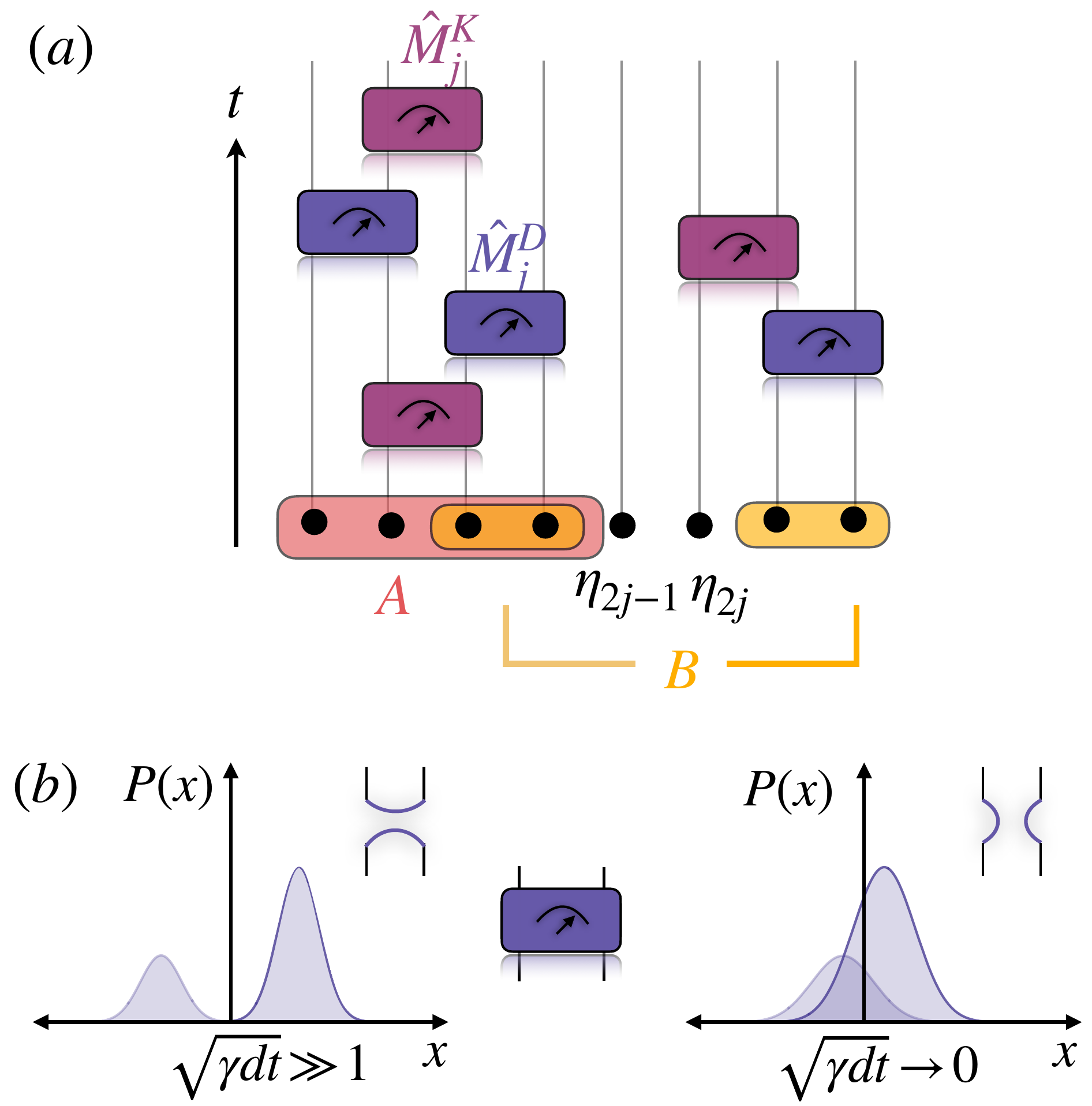}
\caption{(a) Space-time schematic of measurements $\hat{M}^D$, $\hat{M}^K$ on the Majorana (free fermionic) chain with probability $p^D$, $p^K$ and strength $\gamma^D$, $\gamma^K$, respectively. The two partitions of chains A and B contribute to calculating topological entanglement entropy ($S_T$). (b) The probability distribution of the detector outcome $ x$  is modified by the coupling to the system. The modified probability distribution $ P(x)$  given by Eq.(\eqref{distribution}) depends on the coupling strength ($\sqrt{\gamma dt}$).}
\label{fig_model}
\end{figure}
However, how the criticality depends on the detailed coupling to the environment beyond these two limits is yet to be explored. 
\begin{figure*}
\includegraphics[scale=2.2]{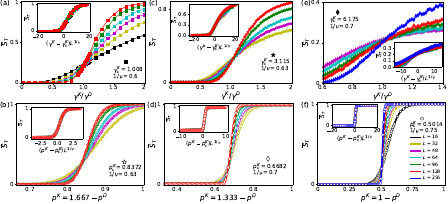}
\caption{The averaged topological entanglement entropy ($\bar{S}_T$) in units of $\log 2$ as a function of (a,c,e) relative measurement strength ($\gamma^K/\gamma^D$) and (b,d,f) probabilities ($p^K,\;p^D$) of the two competing measurements $\hat{M}^K$ and $\hat{M}^D$ for different system sizes. All the inset shows the scaling across the critical points ($\gamma^K_c$ or $p^K_c$) with the corresponding scaling exponents ($\nu$) shown in Fig.~\ref{fig_intro} of the respective figures. The details of different parameters for all cases follow are: ($\blacksquare$) $\gamma^D=1$, $p^D=p^K=1$, ($\bigstar$) $\gamma^D=\gamma^K=3$, $p^K+p^D=1.667$, (\text{\ding{73}}) $\gamma^D=3$, $p^D=p^K=0.833$, ($\Diamondblack$) $\gamma^D=\gamma^K=6$, $p^K+p^D=1.333$, ($\Diamond$) $\gamma^D=6$, $p^D=p^K=0.667$, ($\circ$) $\gamma^D=\gamma^K=9$, $p^K+p^D=1$. The time steps for different cases are (a) $dt=0.05$, (b-f) $dt=1$.}
\label{fig_result}
\end{figure*}

To address this question, we explore the evolution of topological order within a free fermion system driven by measurement-only dynamics, where the system's Hamiltonian is strictly zero, and dynamics arise solely from measurements.
 Free fermionic systems (which preserve the Gaussianity of the state) provide a useful setting, due to their analytical traceability and the efficiency of numerical methods.
We consider two competing measurements: one that creates entanglement between neighboring sites and another that disentangles a local site from the rest of the system, thereby giving rise to distinct area-law phases.
We illustrate that the previously studied projective and weak continuous monitoring can manifest as limiting behaviors within a unified framework of general positive operator-valued measurements (POVM) with variable system-detector coupling. Importantly, this general POVM can be implemented within a framework of a free fermion chain. 

Our findings reveal that the percolation universality that describes the transition in the projective limit for free fermions is unstable to reducing the coupling between the system and detector. In particular, adjusting the coupling strength continuously modifies the finite-size scaling exponent of the transition. 

\section{Model and methodology}
In our study, we consider a free fermionic model 
where measurement-induced entanglement transitions have been explored in monitored random circuits and continuous Hamiltonians with either projective
~\cite{skinner_measurement-induced_2019,PhysRevB.99.224307,PhysRevB.98.205136,fisher_random_2023,PhysRevX.13.041046,PhysRevX.13.041028,klocke2024entanglementdynamicsmonitoredkitaev,PhysRevB.107.245132,PhysRevLett.132.110403,PhysRevB.105.094303,jian2023measurement,PhysRevResearch.6.023176,starchl2024generalized}, or  weak  continuous measurement~\cite{cao_entanglement_2019,szyniszewski_universality_2020,PhysRevLett.126.170602,PhysRevB.103.224210,Biella2021manybodyquantumzeno,PRXQuantum.5.030329,PhysRevB.109.L060302,PhysRevX.13.041045,PhysRevB.110.054313,PhysRevB.111.064313,PhysRevB.110.L060202,PhysRevResearch.6.043246,PhysRevB.110.024303,PhysRevResearch.6.033220,PhysRevX.11.041004,PhysRevB.106.L220304,PhysRevB.108.165126,leung2023theoryfreefermionsdynamics}. 
However, it is possible to adopt a more general perspective to explore a wider range of environments, whereby measurements are described by POVMs ~\cite{brandt_positive_1999,wiseman_quantum_2009,jacobs_quantum_2014}. 
In particular, one can consider POVMs with a controlled back-action,  in which the system is coupled to a detector through unitary evolution, causing them to become entangled. This is described by
$e^{i\lambda\hat{M}\otimes\hat{p}}|\psi\rangle\otimes|\phi\rangle$, where $|\psi\rangle$ and $|\phi\rangle$ are the initial states of the system and detector, respectively. Here, $\lambda=\sqrt{\gamma dt}$  controls the coupling strength between the measured operator $\hat{M}$ and $\hat{p}$ the momentum conjugate to the detectors position $\hat{x}$, whose dynamics can be tracked by $\hat{\dot{x}}=\lambda \hat{M}$~\cite{szyniszewski_entanglement_2019,szyniszewski_universality_2020}. 
A projective measurement of the detector induces a back-action on the system. If the coupling is strong, this is akin to a projective measurement of the system's state. Conversely, under weak coupling, the detector acquires minimal information about the system's state, resulting in a weak back-action that depends nonlinearly on the state itself. 

 The back action on the state depends on the coupling strength as well as the initial preparation of the detector~\cite{PhysRevB.103.224210}. Here, we consider the detector as a one-dimensional pointer, $x $  prepared in a Gaussian distribution centered around zero
with variance given by $\Delta^2$: $|\phi\rangle=\int dx \sqrt{G_\Delta(x)}|x\rangle$, where $G_\Delta(x) = \exp(-x^2/2\Delta^2)/(\sqrt{2\pi}\Delta)$. This pointer is linearly coupled to a local measurement operator $\hat{M}=\sum_\ell m_\ell\hat{\Pi}_\ell$, where $m_\ell$  represents the $\ell^{th}$ eigenvalue of the operator $\hat{M}$ and $\hat{\Pi}_\ell$ stands for the projector onto the corresponding eigenvector. 
 When the detector is projectively measured, the pre-measurement state of the system $|\psi\rangle$ is updated to $|\psi^\prime\rangle$ via~\cite{wiseman_quantum_2009,jacobs_quantum_2014}:
 \begin{align}
|\psi^\prime\rangle=\frac{\hat{K}(x,\lambda,\Delta)}{\sqrt{P(x)}}|\psi\rangle,
\label{kraus_state}
\end{align} 
 where $ \hat{K}(x,\lambda,\Delta)=\sum_\ell\sqrt{G_\Delta(x-\lambda m_\ell)}\hat{\Pi}_\ell$ is the Kraus operator associated with the outcome $x$, of the position measurement on the detector.
The normalization is given by the probability of obtaining the outcome $x$ from the measurement:
\begin{align}
P(x)=\langle\psi|\hat{K}^\dagger(x,\lambda,\Delta)\hat{K}(x,\lambda,\Delta)|\psi\rangle
\label{distribution}.
\end{align}

The choice of coupling strength $\lambda$ allows us to explore the effect of the degree of entanglement between the system and detector on the dynamics. In particular, in the commonly studied weak measurement scenario ($\lambda=\sqrt{\gamma dt}\ll \Delta$), the separation between the Gaussian distributions in Eq.~\eqref{distribution} is small, as shown in right-side of Fig.~\ref{fig_model}(b) and can be approximated by a single Gaussian distribution with a shifted mean value which depends on the state $G_\Delta(x-\lambda \langle \hat{M}\rangle)$. In this scenario, the back action resulting from the measurement is  akin to a Wiener process~\cite{jacobs_quantum_2014} and produces dynamics equivalent to the Stochastic Schr\"{o}dinger's equation~\cite{barchielli_quantum_2009,wiseman_quantum_2009,cao_entanglement_2019}, as given by:
\begin{align}
\label{wiener}
|\psi^\prime\rangle\approx 
\frac{1}{N}e^{(\delta W_t+\lambda^2\langle\hat{M}\rangle)\hat{M}  }|\psi\rangle,
\end{align}
where the Wiener stochastic increments are independently Gaussian-distributed with $\langle\delta W_t\rangle=0$ and $\langle \delta W_t \delta W_{t'}\rangle=\lambda^2 \delta_{t,t'}$ \footnote{ Instead if the detector is a two-level system prepared in the ground state $ |Z_-\rangle_d$, the back action on the state would result in the quantum jump equation, see discussion in Ref. \cite{PhysRevB.103.224210}}. 
Conversely, in the case of strong measurement ($\lambda\gg\Delta$), the distribution of detector outcomes consists of a sum of separated Gaussians with exponentially small overlap, 
as illustrated on the left side of Fig.~\ref{fig_model}(b). Consequently, the outcome of the detector feedback on the system is dominated by a one eigenvector projection over all others of the measurement $\hat{M}$. The resulting dynamics are exponentially close to those of a projective measurement. 

The above discussion is general and applicable to any many-body system subjected to the monitoring of local operators. However, implementing a weighted sum of projectors numerically is typically challenging. Nonetheless, numerical tractability can be achieved in a free fermion (i.e., Gaussian \footnote{Gaussian quantum states refer to eigenstates of quadratic Hamiltonians. These are states that can be described as a Slater determinant and are fully characterized by the two-point correlation matrices of the form $C_{ij}(t)  =\langle c_i^\dag c_j \rangle $ and $F_{ij} =\langle c_i^\dag c_j^\dag\rangle $.}) state evolving under weak continuous~\cite{cao_entanglement_2019,PhysRevLett.126.170602,kells_topological_2023,PhysRevB.108.165126,PhysRevB.107.064303,Biella2021manybodyquantumzeno,PhysRevB.103.224210} or in Clifford gates subjected to projective monitoring~\cite{PhysRevB.99.224307,lavasani_measurement-induced_2021,lavasani_topological_2021,Quantuminformation,skinner_measurement-induced_2019}. It's worth noting that measurements satisfying $\hat{M}^2=\hat{M}$ or $\hat{M}^2=\hat{\mathbb{I}}$ preserve the gaussianity of the state (i.e. Slater determinant form).  Hence, in free fermions, it's possible to express the Kraus operator for a {\it general} system and detector coupling strength as:
\begin{eqnarray}\label{eq:general update}
  |\psi_{t+dt}\rangle  &=&\frac{1}{N} e^{ \frac{\lambda P^M x\hat{M}}{2\Delta^2} } |\psi_{t}\rangle,
\end{eqnarray}
where we have added a probability $P^M\in\{0,1\} $ to perform the given measurement, which drives the transition in the projective limit (appendix~\ref{appx:1}). 

With a comprehensive understanding of POVMs, we now focus on our system of interest. We study measurement-only dynamics driven by two non-commuting POVMs on a one-dimensional spinless free fermionic chain (Majorana string)  of length $L$ as shown in Fig.~\ref{fig_model}(a). 
The two types of measurements correspond to local density and Kitaev-type bond density and can be expressed as staggered pairwise parity checks on the Majorana fermions:
\begin{align}\label{eq:parityMeasurements}
\hat{M}^D_j&=c^\dagger_jc_j-c_jc^\dagger_j=i\eta_{2j}\eta_{2j-1},\\
\hat{M}^K_j&=\left(c_j+c_j^\dag\right)\left(c_{j+1}-c_{j+1}^\dag\right)
=i\eta_{2j}\eta_{2j+1},
\end{align}
where $c_j= \frac{1}{2}\left(\eta_{2j}+i\eta_{2j-1}\right)$ is fermionic annihilation  operator on the $j^{th}$ site.
The measurement strength is controlled by parameters $\gamma^D$ and $\gamma^K $. The eigenvalues of each measurement are $1,-1$ with the corresponding projectors given by $\hat{\Pi}^D_{j,\pm}= (1\pm i \eta_{2j}\eta_{2j-1})/2$, $\hat{\Pi}^K_{j,\pm}= (1\pm i \eta_{2j}\eta_{2j+1})/2$ ,  for the density and Kitaev-type measurement, respectively.\\

The chain evolves in discrete time steps $dt$. During each time step $dt$, the update runs over all chain sites, performing two measurement operations with probabilities $p^K$ and $p^D$, respectively.  
Following  Eq.~\eqref{eq:general update}, the total update to the state after each time step $dt$ is given by:
\begin{align}\label{eq:state_update}
|\psi_{t+dt}\rangle=\displaystyle\prod_{j=1}^L\frac{e^{\frac{P^K_j\sqrt{\gamma^K dt}  x^K_j\hat{M}^K_j}{2\Delta^2}}}{N^K_j}\frac{e^{\frac{P^D_j\sqrt{\gamma^D dt} x^D_j \hat{M}^D_j}{2\Delta^2}}}{N^D_j} |\psi_t\rangle,
\end{align}
where $\hat{M}^{K,D}_j$ are 
given in eq. \eqref{eq:parityMeasurements} with measurement strength $\gamma^{K,D}$, $P^{K,D}_j$ are binary variables, indicating whether to monitor (1) or not monitor (0) the site $j$ of the system with the corresponding measurements (for OBC $P^{K}_L$ is set to $0$). Here, $\Delta$ is the width of the initial Gaussian state of the detector/pointer, and $x^{K,D}_j$ is the readout of the pointer used as feedback to the system. The state is normalized after each single measurement with $N^{K,D}_j$; therefore, the dynamics are both stochastic and non-linear. 
In the weak coupling limit $\gamma^{K,D} dt \rightarrow 0 $ and for $p^K=p^D =1$ ($P^{K,D}_j=1\forall j$), we recover the stochastic Schrodinger equation~\cite{kells_topological_2023}, Eq. \eqref{wiener}. Conversely, for very strong fixed coupling $\gamma^{K,D} dt \gg1 $ and $p^K+p^D =1$ ($P^{K,D}_j=0,1\forall j$), the dynamics are exponentially close to projective measurements~\cite{lavasani_topological_2021}. 

Under repeated measurements, the average density matrix evolves toward the maximally mixed state as its long-term fixed point. Capturing non-trivial effects from measurements requires tracking averages of quantities that are non-linear in the density matrix, such as entanglement entropy.
The competing dynamics drive a measurement-induced phase transition between topologically distinct disentangled phases (see Fig. \ref{fig:biEE}). In particular, Kitaev-type measurements direct the system towards states with symmetry-protected topological order. We characterize this transition using the topological entanglement entropy~\cite{Quantuminformation,kitaev_topological_2006,PhysRevLett.96.110405,zeng_topological_2016,PhysRevB.101.085136,10.21468/SciPostPhysCore.3.2.012,PhysRevB.105.085106}:
\begin{align}
\label{topoEE}
\bar{S}_T=\bar{S}_{A}+\bar{S}_{B}-\bar{S}_{A\cup B}-\bar{S}_{A\cap B},
\end{align}
a combination of von-Neumann entropies, i.e., $S_X=-\text{Tr}(\rho_X\log\rho_X)$ where  $\rho_X$ is the reduced density matrix of subsystem $X$ and bar indicates the average over trajectories. A possible partition into subsystems  A and B  is illustrated in Fig.~\ref{fig_model}(a).
Given the Gaussianity of the dynamics, we can express the density matrix in terms of the correlation matrix, facilitating the calculation of the entanglement entropy~\cite{peschel_calculation_2003,keating_random_2004,peschel_reduced_2009,PhysRevResearch.2.013175}.

Kitaev-type measurements induce short-range entanglement between adjacent sites, resulting in a non-zero sub-leading contribution to the entanglement entropy $S_T=\log 2$, while density measurements disrupt it by projecting onto local density eigenvectors, with $S_T\rightarrow 0$. This insight anticipates a transition from a trivial area law to a topological area law phase (appendix~\ref{appx:1}), governed by the parameters controlling the measurement rates or probabilities. In the following, we use two tuning parameters: the probability of a Kitaev measurement $p^K$ for a fixed sum of probabilities $p^S = p^K + p^D \in [0,2]$, along with the relative measurement strength $\gamma^K/\gamma^D$. Near the transition point, the topological entanglement entropy follows a scaling law~\cite{skinner_measurement-induced_2019}, expressed as:
\begin{align}
\label{sTopoEE}
\bar{S}_T(q,L)=F((q-q_c)L^{1/\nu}),
\end{align}
where $F(x)$ is some universal scaling function, $q_c$ is the critical value of the tuning parameter $q$ across the transition, and $\nu$ is the critical
exponent associated with divergence of length scale 
across the critical point $q_c$, i.e., $\xi\propto |q-q_c|^{-\nu}$ ~\cite{skinner_measurement-induced_2019}.

To quantify the best collapse for the topological entanglement data, we analyze the objective function $\epsilon(\nu)$~\cite{lavasani_measurement-induced_2021}, defined as the mean squared deviation of the topological entanglement entropy for fixed system size, from its optimally fitted value:
\begin{align}
\label{objfxn}
\epsilon(\nu)=\frac{1}{n-2}\displaystyle\sum_{i=2}^{n-1}(y_i-\bar{y}_i)^2,
\end{align}
where $y_i=\bar{S}_T(q_i,L)$,  and $\bar{y}_i-y_{i+1}=\frac{y_{i+1}-y_{i-1}}{x_{i+1}-x_{i-1}}(x_i-x_{i+1})$ is the anticipated or optimal value of $y_i$ with 
 $x_i=(q_i-q_c)L_i^{1/\nu}$.
 Here, $i$ runs over all data points  $q_i=\gamma^K_i$ or $q_i=p^K_i$ across various system sizes,  $L_i$, arranged based on their $x$ values such that $x_1 < x_2 <\hdots < x_n$, with $n$ representing the total number of data points.
The objective function $\epsilon(\nu)$ is then minimized to determine the precise critical exponent ($\nu$) corresponding to a critical point ($\gamma^K_c$ or $p^K_c$). 

In the strong projective limit, dynamics map to a classical 2D percolation model with critical exponent $\nu=4/3$ \cite{lavasani_topological_2021,PhysRevX.13.041028}. However, this mapping doesn't extend to the weak continuous limit, where the dynamical evolution is nonlinear, yielding a critical exponent of $\nu=5/3$ \cite{kells_topological_2023}. Our goal is to utilize the update in Eq. \eqref{eq:state_update} to explore the continuous evolution of the entanglement transition across these universality classes.

\section{Results} 
\begin{figure}[h!]
\centering
\includegraphics[width=.45\textwidth]{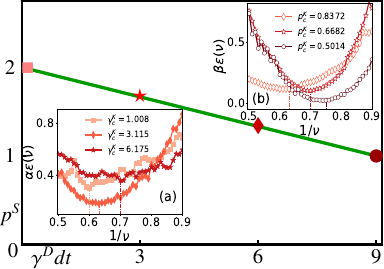}
\caption{Schematic of the various critical exponents ($\nu$) and corresponding critical points ($\gamma^K_c$, $p^K_c$) obtained along the line ($9p^S+\gamma^Ddt=18$) in the frame of total probability of measurements ($p^S=p^K+p^D$) versus effective density measurement strength ($\gamma^Ddt$). The data for different system sizes is collected in two ways: (a) $\gamma^Ddt$, $p^K=p^D=p^S/2$, and $\gamma^Kdt$ are varied. (b) $\gamma^Ddt=\gamma^Kdt$, and $p^K,\;p^D$ are varied by following the relation $p^K+p^D=p^S$.  The other specific choice of parameters for different cases is as follows: ($\blacksquare$) $dt=0.05$, $\gamma^D=1$, $\alpha=14000$, ($\bigstar$)  $dt=1$, $\alpha=2500$, $\beta=5000$, ($\Diamondblack$) $dt=1$, $\alpha=12000$, $\beta=2500$, ($\medbullet$) $dt=1$, $\beta=300$.}
\label{fig_intro}
\end{figure}
As a first step towards tracking the universal properties of the transition for generic system and detector coupling, we consider the two previously studied limits of weak continuous monitoring ($\gamma dt\rightarrow 1 $, and $p^K=p^D =1 $) and strong projective measurement ($\gamma dt \gg1 $, and $p^K+p^D =1 $) for system sizes ranging from $L=16$ to $L=256$. 

Fig.~\ref{fig_result}(a) shows the finite-size scaling analysis of the topological entanglement entropy as a function of the tuning parameter $\gamma^K/\gamma^D$ for the weak continuous limit. The analysis is performed with small time steps $dt=0.05$; such that the effective measurement strengths are weak, i.e., $\gamma^{K,D}dt\to 0$. Thus, the back-action from the detector mirrors the Wiener process and results in stochastic Schr\"{o}dinger dynamics in our controlled back-action setup. 
The inset of Figure~\ref{fig_result}(a) presents the data collapse around the critical point $\gamma_c^K/\gamma^D\approx1.008(4)$ (obtained by averaging the crossing of different system sizes)
 following Eq.~\eqref{sTopoEE} with the critical exponent $\nu=5/3$ obtained using  Eq.~\eqref{objfxn}.

To study the best fit to the scaling exponent, we examine the objective function $\epsilon(\nu)$ defined in Eq.~\eqref{objfxn} from the optimally fitted data of TEE (Appendix~\ref{appx:3}).
The results are depicted as pink squares in Figure~\ref{fig_intro}(a).
We find that the critical exponent which minimized $\epsilon(\nu)$ is in agreement ($\gamma^K_c=\gamma^D$, $\nu=5/3$) with previous results on weak continuous monitoring of free fermions~\cite{kells_topological_2023}. 
Next, we examine the strong coupling limit, which corresponds to performing a projective measurement on the system's local degrees of freedom.   
The topological entanglement undergoes a phase transition at a critical probability value of $p^K_c= 0.501(4)$ in the system, as shown in  Fig.~\ref{fig_result}(f). The corresponding data collapse onto a universal scaling function with critical exponent $\nu=4/3$ is shown as brown circles in the inset of Fig.~\ref{fig_intro}(b). Another verification for the critical exponent is provided by the study of the objective function, showing a minimum at the critical exponent $\nu=4/3$ for a critical value $p_c^K=0.501(4)$ as depicted in Fig.~\ref{fig_intro}(b). This critical exponent value precisely matches the percolation prediction~\cite{lavasani_measurement-induced_2021}. 

Having established that our methodology produced the correct scaling in the two limits, we next tackled the evolution of the universal behavior between them.  For this purpose, we follow a line in the parameter space spanned by $p^S $ and $\gamma^D dt $, which connects the projective limit, marked by $\medbullet $, 
and the continuous limit, marked by $\blacksquare $, 
 and examine the critical properties of the topological entanglement transition
at intermediate points along the line, see Fig.  \ref{fig_intro}. 


We present the results of our study for two intermediate points marked by a $\bigstar$  ($p^S=1.667$, $\gamma^Ddt=3$) and a $\Diamondblack $ ($p^S=1.333$, $\gamma^Ddt=6$)  in Fig.~\ref{fig_intro}.
For each point of these points in the parameter space, we tune across the transition in two ways:\\
(i) Fixing  equal measurement probabilities $p^D=p^K=p^S/2$, while varying the relative  measurement strength ($\gamma^K/\gamma^D$) .\\
(ii) Fixing equal measurement strength $\gamma^K=\gamma^D$ while varying the relative probabilities of the two measurements such that their sum is fixed, i.e., $p^K = p^S-p^D$. 

The finite size scaling of the topological entanglement entropy (TEE) for the  $\bigstar$ and for the $\Diamondblack $  in Fig. \ref{fig_intro} are shown in Fig. \ref{fig_result} (b)-(c), and (d)-(e), respectively. Data collapses near the critical point are illustrated in the inset of the respective figures and confirmed by the objective function $\epsilon(\nu)$ in Fig.~\ref{fig_intro} (a,b) plotted for star and diamond markers.
Remarkably, for each of the intermediate points ($\bigstar$, $\Diamondblack $), the same critical exponent is obtained when tuning across the transition using (i) the relative strength ($\gamma^K$/$\gamma^D$) as $\bigstar,\Diamondblack$, Fig.~\ref{fig_intro}(a) or  (ii) the relative probabilities $p^K = p^S-p^D$ as  \text{\ding{73}}, $\Diamond$ in Fig.~\ref{fig_intro}(b).
Notably, the critical exponent monotonously increases between the weak continuous ($\blacksquare$) and the projective ($ \medbullet$) configurations. 

\begin{figure}[h!]
\centering
\includegraphics[width =0.4\textwidth]{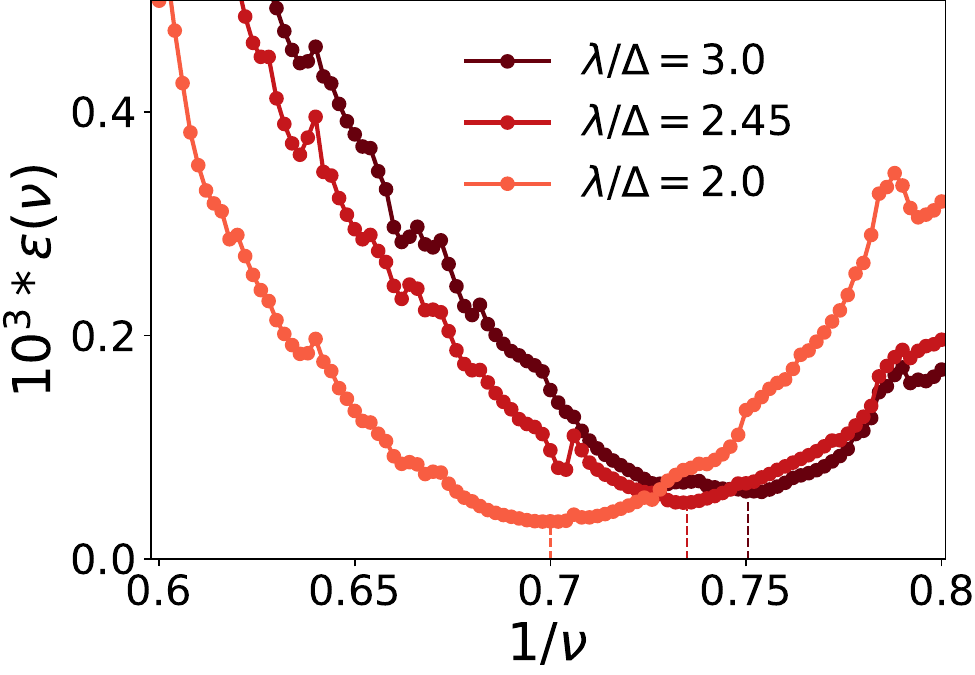}
\caption{The objective function ($\epsilon(\nu)$) as a function of critical exponents ($\nu$) corresponding to different separations between two Gaussian distributions of the detector's state ($2\lambda/\Delta$). Other parameter selections include: $dt=\Delta=1$, $\gamma^D=\gamma^K=4,6,9$, $\lambda=\sqrt{\gamma^{K}dt}=\sqrt{\gamma^{D}dt}$ and $p^K=1-p^D$ , considering various system sizes.}
\label{fig_obj}
\end{figure}

Finally, we study the stability of the percolation universality with the reduction of entanglement between the system and the detector. For this purpose, we study the correlation length critical exponent by perturbing around the strong coupling limit by reducing the coupling strength $\lambda $, keeping the ratio $\gamma^K/\gamma^D $ and the sum of measurement probabilities $p^S $ fixed to $1 $. This corresponds to reducing the separation between the two Gaussian distributions ($\lambda/\Delta$), and is equivalent to moving along the horizontal line passing through the point $ \medbullet$ in  Fig.~\ref{fig_intro}. We find a significant change in the scaling exponent of the TEE phase transition, $\nu$, even in the regime $\lambda/\Delta>1$. This finite size analysis indicates that the projective limit is destabilized as the coupling between the system and detector is weakened. This is exemplified by the shift of the global minimum of the objective function for different values of $\lambda/\Delta$ in Fig.~\ref{fig_obj}. 
Nevertheless, the possibility of a sharp transition in the thermodynamic limit cannot be excluded based on numerical analysis.


\section{Discussion and conclusion} We developed a general methodology to study the dynamics driven by competing Gaussian measurements, with tunable coupling between the system and detector.
This method continuously interpolates between projective measurement and weak continuous monitoring. In the strong projective limit, the dynamics can be mapped onto 2D percolation~\cite{lavasani_measurement-induced_2021,10.1063/1.1704215,PhysRevE.73.066116,roosz_nonequilibrium_2016,stauffer2018introduction} and the critical properties are confirmed numerically. Meanwhile, weak non-projective measurements exhibit different universal behaviors~\cite {kells_topological_2023}. Our work establishes a crucial link between these limits and allows to track the evolution of the universal properties of the topological entanglement transitions between these two limiting behaviors. 

We find that the critical exponents that characterize the diverging correlation length at the transition grow monotonously as one moves away from the projective limit $\nu =4/3 $ towards the weak continuous limit $\nu=5/3 $. The monotonous increase in the correlation length $(p-p_c)^{-\nu} $ with the reduced measurement strength can be understood from the percolation picture when perturbing around the strong projective limit. Here, measurements are mapped to cuts in the network \cite{lavasani_topological_2021}. When perturbing away from this limit by weakening the measurement strength, this would partially ``heal" a fraction of cuts in the grid, thus increasing the correlation length of a connected cluster. Our finite-size analysis indicates that the mapping to 2D classical percolation fails to describe the generic coupling between system and environment, even when that coupling is relatively strong.

\section{Acknowledgments} We thank Graham Kells and  Chun Y. Leung for their valuable input. A.R. acknowledges support from the Royal Society, grant no. IECR2212041. 
R. N. acknowledges the financial support of the Krietman fellowship at BGU and thanks the ICTP Visitors Program for  hospitality. D. M. thanks the GMT group at SPEC CEA and LPTMS CNRS for their hospitality. The authors
thank the Institut Henri Poincaré (UAR 839 CNRS-Sorbonne Université) and the LabEx CARMIN (ANR-10-LABX-59-01) for their support and the cluster facilities at Ben-Gurion University, Lancaster University, and Laboratoire de Physique des Solides CNRS.

After the completion of this manuscript, we became aware of related work exploring the role of different couplings to ancilla detectors~\cite{aziz2024critical,PhysRevB.109.224203}.

\onecolumn
\section*{Appendix}
\appendix

\section{\label{appx:1}Free-fermionic methods and numerical techniques}
In this appendix, we present the free-fermionic numerical techniques used in this work. The dynamic evolution considered in this work preserves the Gaussianity of the state. Consequently, the quantum state can be represented as a Slater determinant at any time $t$ within the Bogoliubov de Gennes (BdG) formalism, as the measurements are not number-conserving.
\begin{align}
|\psi_t\rangle:=
\begin{bmatrix}
U^t \\
V^t \end{bmatrix}_{2L\times N_p}\hspace{-0.3cm}=\prod_{n=1}^{N_p}\big[\sum_{i=1}^{L}(U^t_{i,n}c^\dagger_i+V^t_{i,n}c_i)\big]|0\rangle,
\end{align}
where $|0\rangle$ is the state with all c-fermion sites of the chain left unoccupied and $N_p=L$ is the number of particles. The state is updated according to Eq.~\eqref{eq:state_update}, namely:
\begin{align}
|\psi_{t+dt}\rangle=\displaystyle\prod_{j=1}^L\frac{1}{N^K_j}e^{\frac{P^K_j\sqrt{\gamma^K dt}  x^K_j\hat{M}^K_j}{2\Delta^2}}\frac{1}{N^D_j}e^{\frac{P^D_j\sqrt{\gamma^D dt} x^D_j \hat{M}^D_j}{2\Delta^2}}|\psi_t\rangle,
\end{align}
We note that the update involves two random numbers: (i) The probability to perform a given measurement at site $j$, $P_j^{K/D}$ is sampled from a binomial distribution with a success rate set by the global measurement probability   $p^{K/D}$. (ii) The random number $x_j^{K/D}$ is the readout of the pointer and whose probability distribution follows the Born rule $P(x_j^{K/D}) = \langle \psi_{t} |\hat{K}_j^\dag(x_j^{K/D},\gamma^{K/D}) \hat{K}_j(x_j^{K/D},\gamma^{K/D})|\psi_{t} \rangle $. 

The updated state $|\psi_{t+dt}\rangle$ is used to compute the entanglement entropy. 
For a Gaussian state, Wick’s theorem simplifies this computation significantly by reducing the problem from diagonalizing the full  $2^L\times2^L$ density matrix to a simpler problem of finding the eigenvalues of the $2L\times 2L$ correlation matrix~\cite{peschel_reduced_2009,PhysRevB.101.085136}. The construction of the correlation matrix in the BdG formalism follows:
\begin{align}
    \mathcal{C}_t=\begin{bmatrix}
    \langle c_t^\dagger c_t\rangle &\langle c_t^\dagger c^\dagger_t\rangle\\
    \langle c_tc_t^\rangle &\langle c_t c_t^\dagger\rangle
    \end{bmatrix}=\begin{bmatrix}
    U_tU_t^\dagger &U_tV_t^\dagger\\
    V_tU_t^\dagger &V_tV_t^\dagger
    \end{bmatrix}_{2L\times 2L}
\end{align}
The entanglement entropy of any partition is then given by
\begin{align}
S_\ell=\frac{1}{2}\sum_{i=1}^{2\ell} (-\lambda_i \log(\lambda_i)-(1-\lambda_i) \log(1-\lambda_i)),
\end{align}
where $\lambda_i$ are the eigenvalues of the reduced correlation matrix corresponding to the chosen partition. 
\begin{figure}[h!]
    \centering
    \includegraphics[width=0.4\linewidth,height=0.265\linewidth]{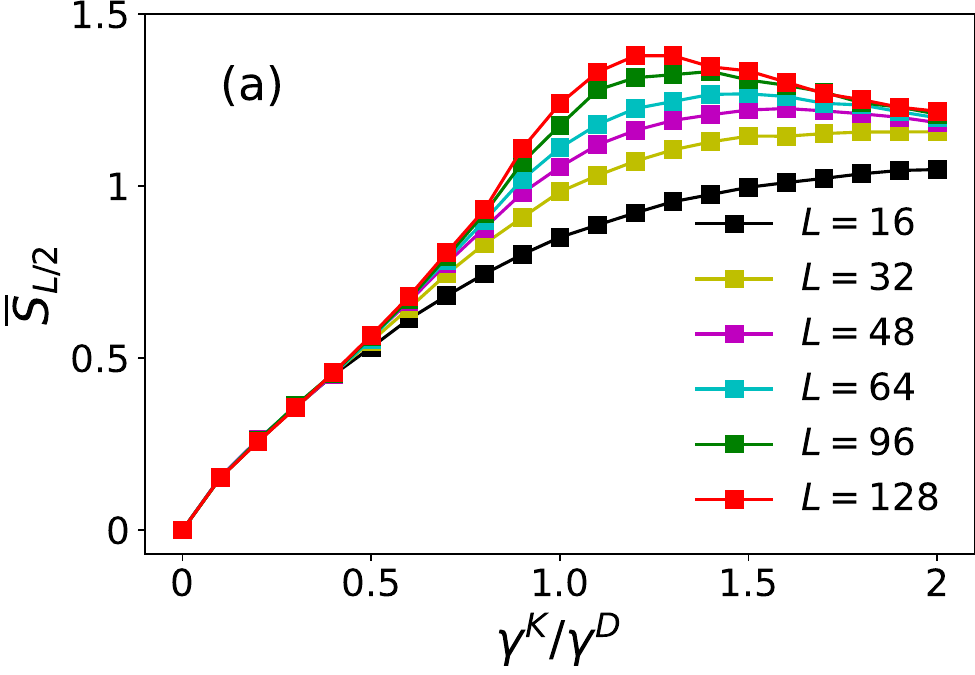}
\includegraphics[width=0.4\linewidth,height=0.265\linewidth]{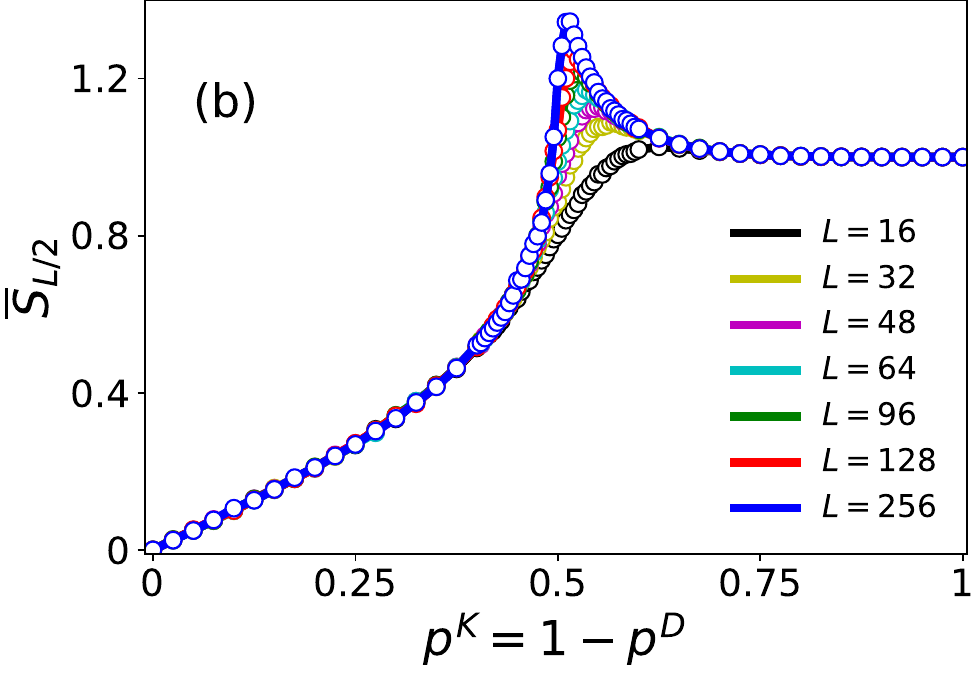}
    \caption{Averaged bipartite entanglement entropy ($\bar{S}_{L/2}$) in units of $\log(2)$ as function of (a) the relative measurement strengths $(\gamma^K/\gamma^D)$, with both measurement probabilities fixed at unity ($p^K = p^D = 1$) and time step $dt = 0.05$, and (b) the measurement probabilities $(p^K = 1 - p^D)$, with fixed strengths $\gamma^K = \gamma^D = 9$ and time step $dt = 1$, for various system sizes.}
    \label{fig:biEE}
\end{figure}
The reduced correlation matrix allows us to study the entanglement of various partitions of the system.  Fig.~\ref{fig:biEE} shows the bipartite entanglement entropy ($\bar{S}_{L/2}$), in the steady state, averaged over a few hundred trajectories for different system sizes. Notably, $\bar{S}_{L/2}$ remains independent of system size in both phases away from the transition point, indicating a measurement-induced phase transition between distinct area-law~\cite{eisert_colloquium_2010,wolf_area_2008} regimes.

In the continuous limit, measurements are performed everywhere but with weak strength, meaning $p^{K/D}=1$, such that  $P^{K/D}_j=1\forall j$, and the phase transition is governed by the relative measurement strengths $ \gamma^{K}/\gamma^{D}$. 
\begin{figure}[h!]
\centering
\includegraphics[scale=0.33]{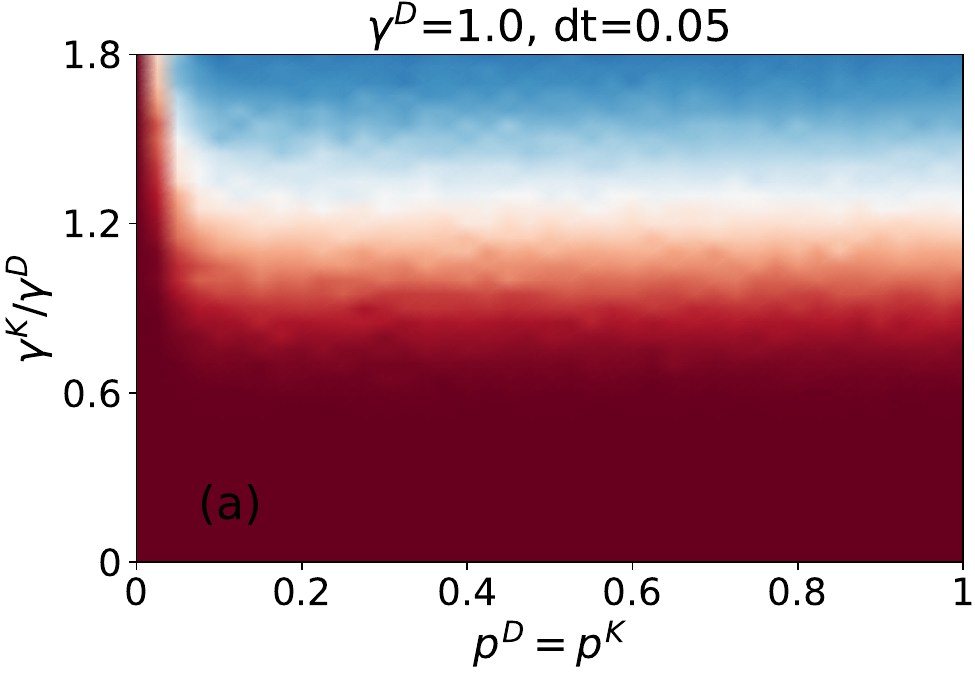} 
\includegraphics[scale=0.33]{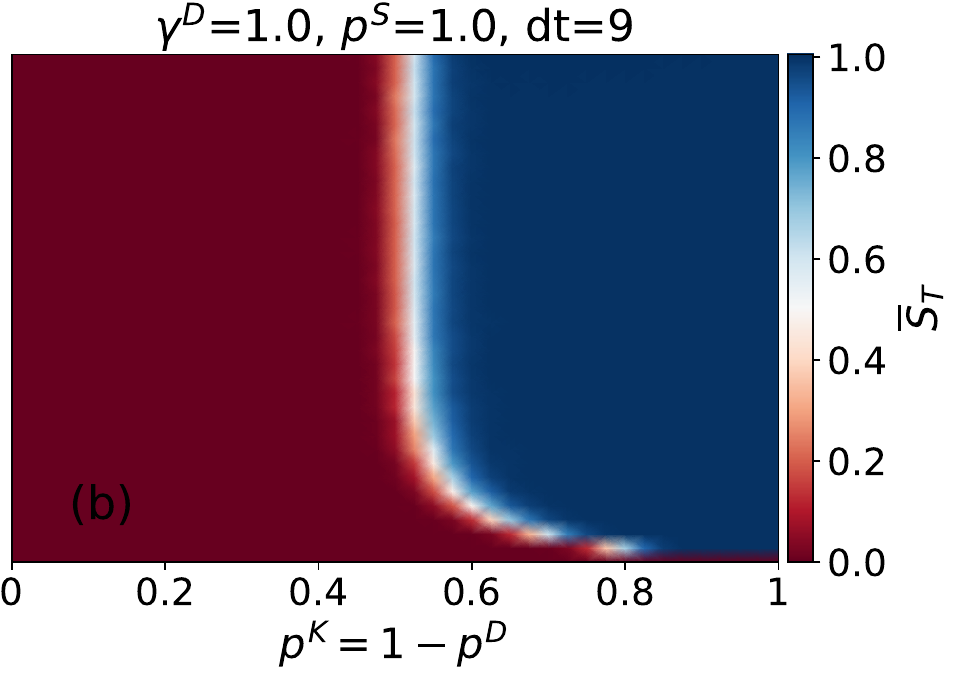}
\includegraphics[scale=0.33]{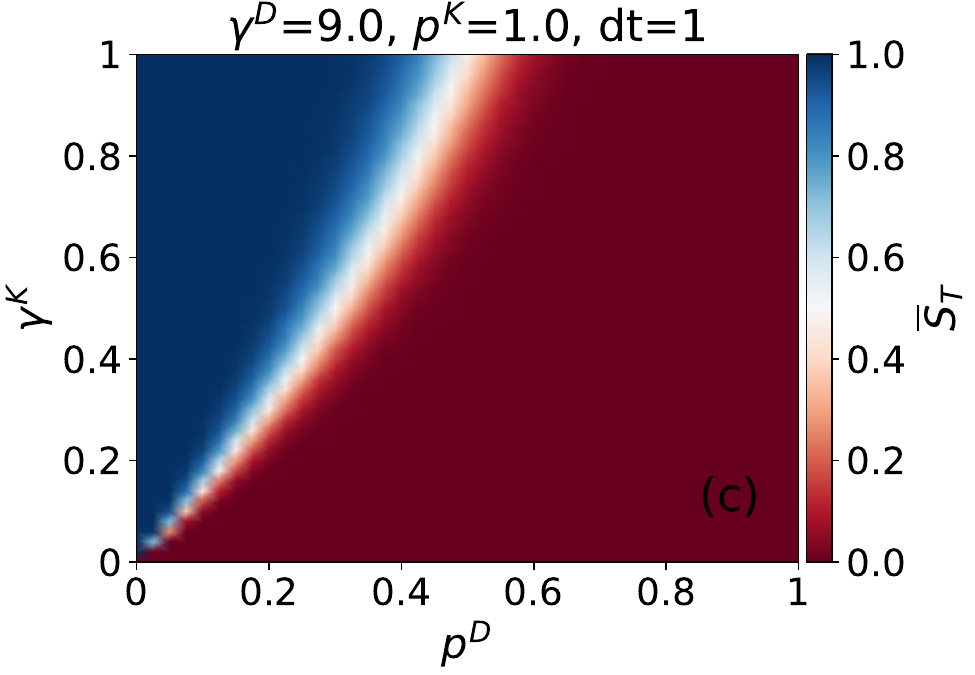}
\caption{The averaged topological entanglement entropy ($\bar{S}_T$) for $L=64$ as a function of measurement strength ($\gamma^K$) and probability ($p^D$). The choice of parameters for different cases is as follows: (a) $\gamma^D=1$, $p^D=p^K$ with $dt=0.05$, (b) $\gamma^D=1$, $p^K+p_D=p^S=1$ with $dt=9$, (c) $\gamma^D=9$, $p^K=1$ with $dt=1$.}
\label{fig:surf}
\end{figure}
In contrast, for projective measurements, both measurement strengths $\gamma^{K/D} $ are strong and fixed, making the phase transition primarily dependent on the likelihood of a particular measurement occurring, which is governed by the probability $p^{K/D}$. This behavior is shown in  Fig. \ref{fig:surf} for a fixed system size $L=64$ with various measurement parameters. 

The choice of short time step $dt=0.05$  in Fig.~\ref{fig:surf}(a) corresponds to weak continuous measurement, i.e., $\sqrt{\gamma^{K,D}dt}<1$. Here, the transition occurs at $\gamma^K/\gamma^D=1$ after sufficiently high measurement probabilities. The case of large time step $dt=9$ results in a strong measurement strength $\sqrt{\gamma^{K,D}dt}>1$, as illustrated in Fig.~\ref{fig:surf}(b), indicating the strong projective limit. In this case, the transition occurs at $p^K=p^D $ and remains the same after suitably strong measurement strengths roughly $\gamma^Kdt=\gamma^Ddt=9$. Figure~\ref{fig:surf}(c) depicts a mixed scenario where phase transition happens at different Kitaev measurement strengths ($\gamma^K$) with a change in the probability of performing a density-measurement rate $p^D$. 

\section{\label{appx:3}Scaling the fitted data around the critical point $\gamma^K_c$}
\begin{figure}[h!]
\includegraphics[scale=0.3545]{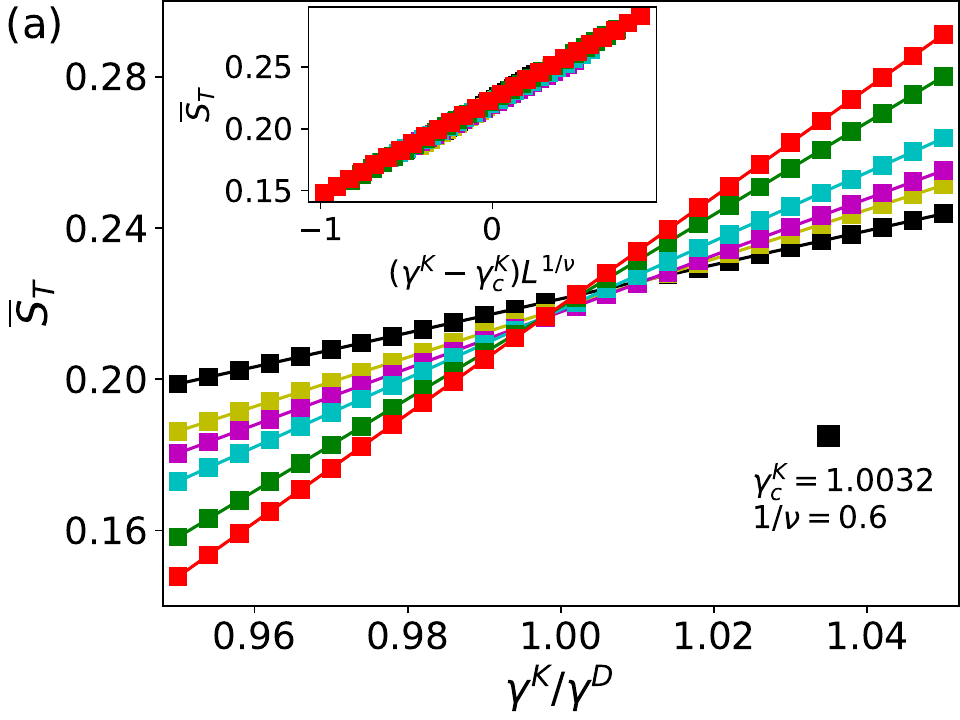}
\includegraphics[scale=0.3575]{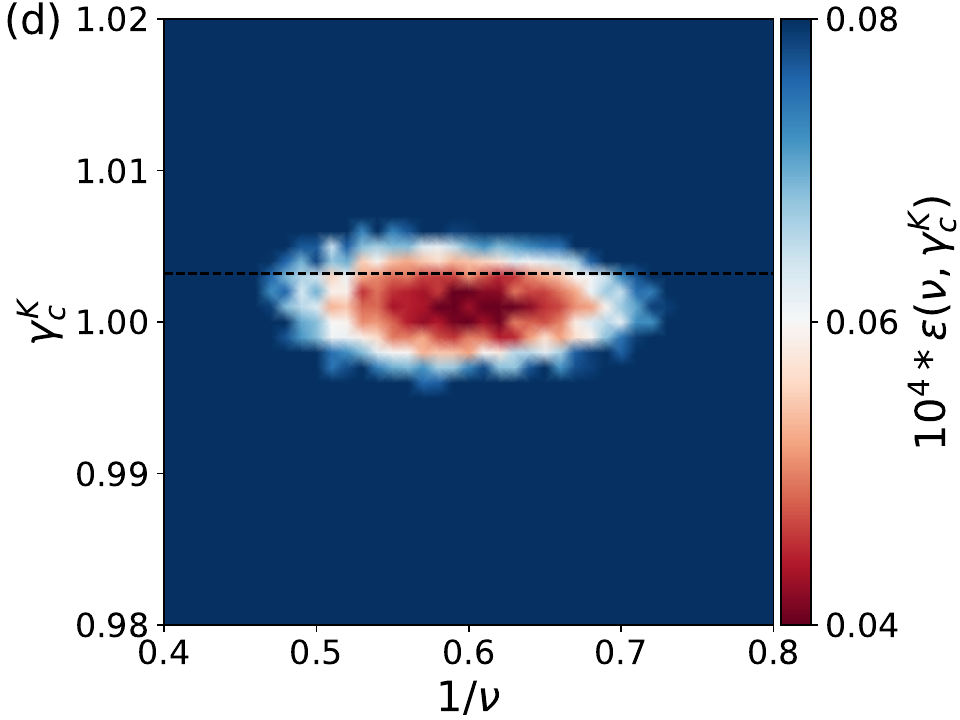}
\includegraphics[scale=0.3545]{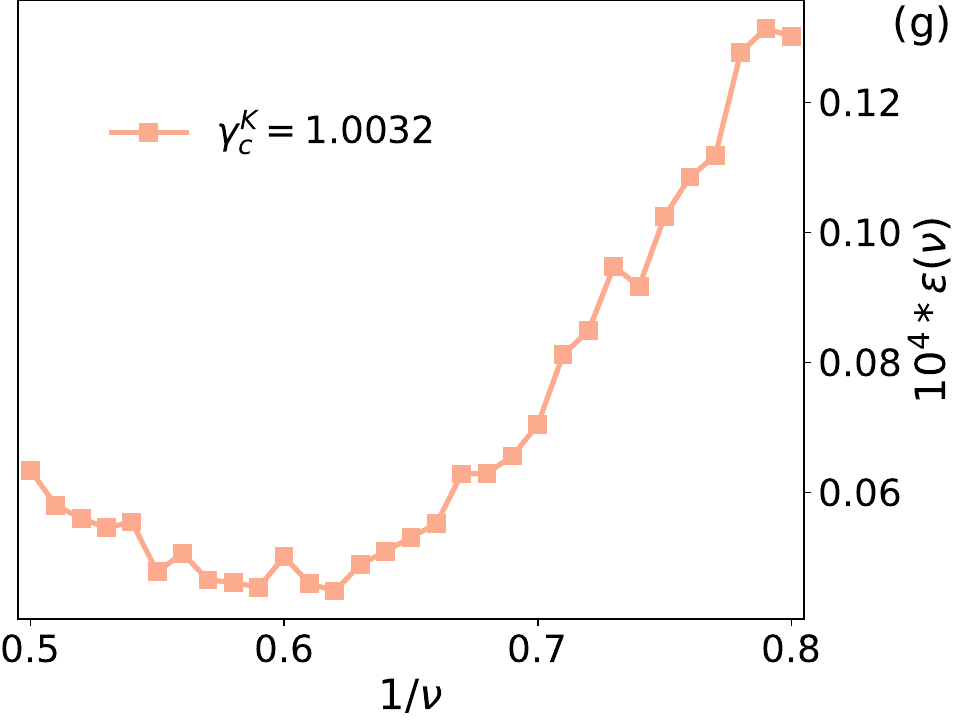}\\
\includegraphics[scale=0.3545]{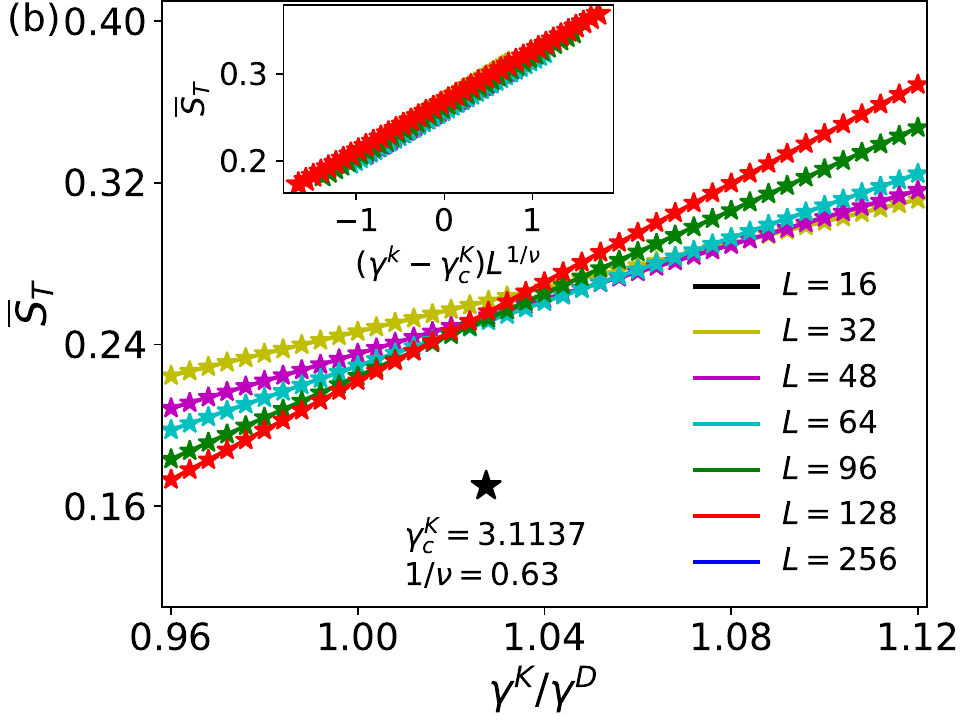}
\includegraphics[scale=0.3575]{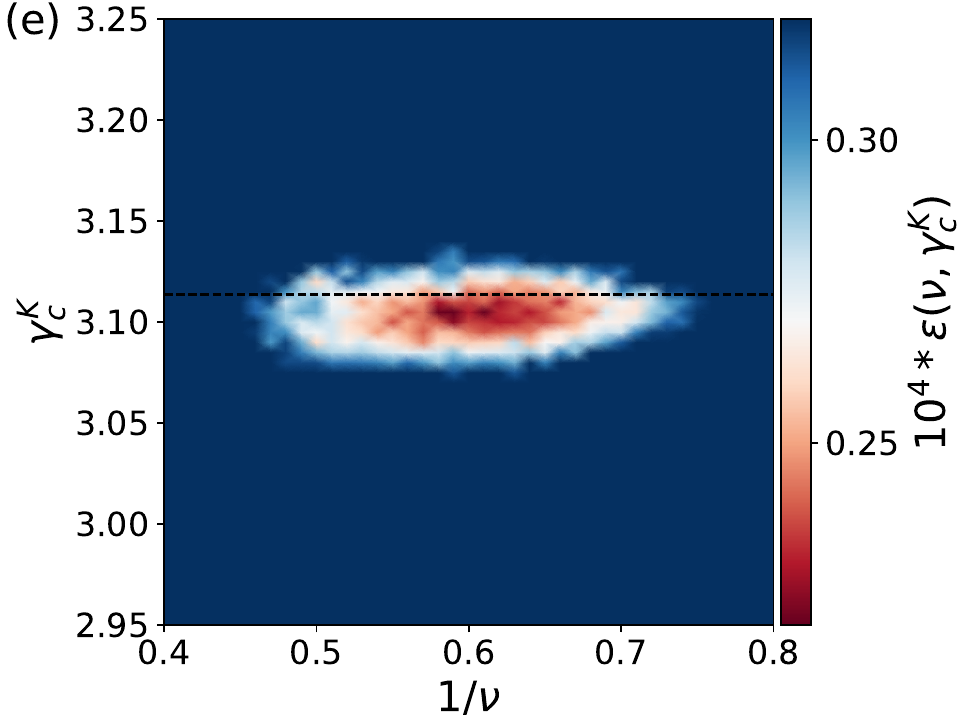}
\includegraphics[scale=0.3545]{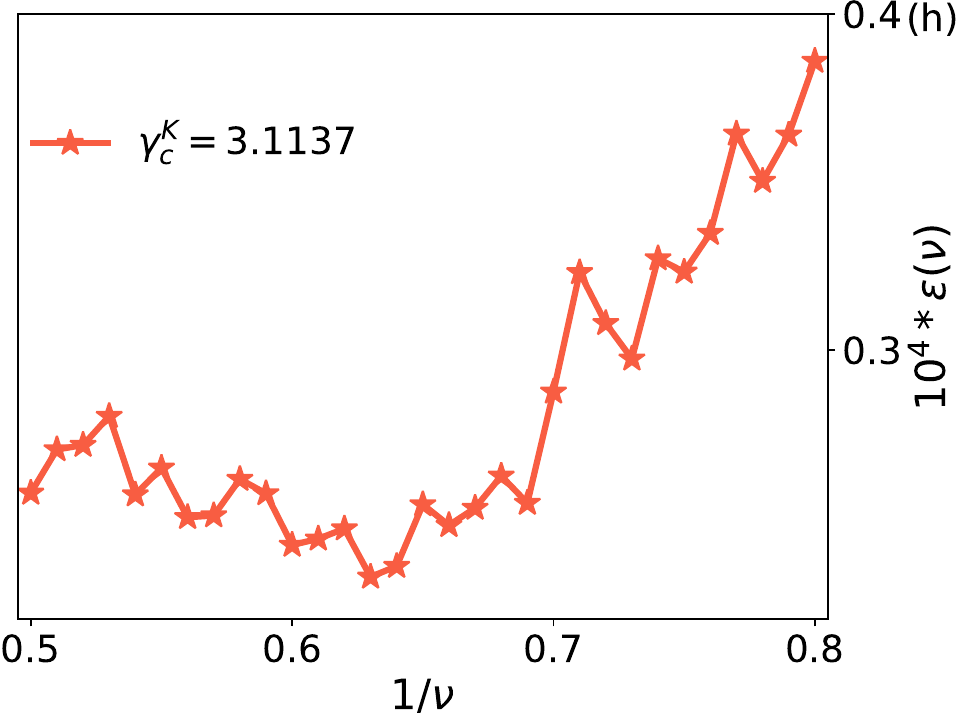}\\
\includegraphics[scale=0.3545]{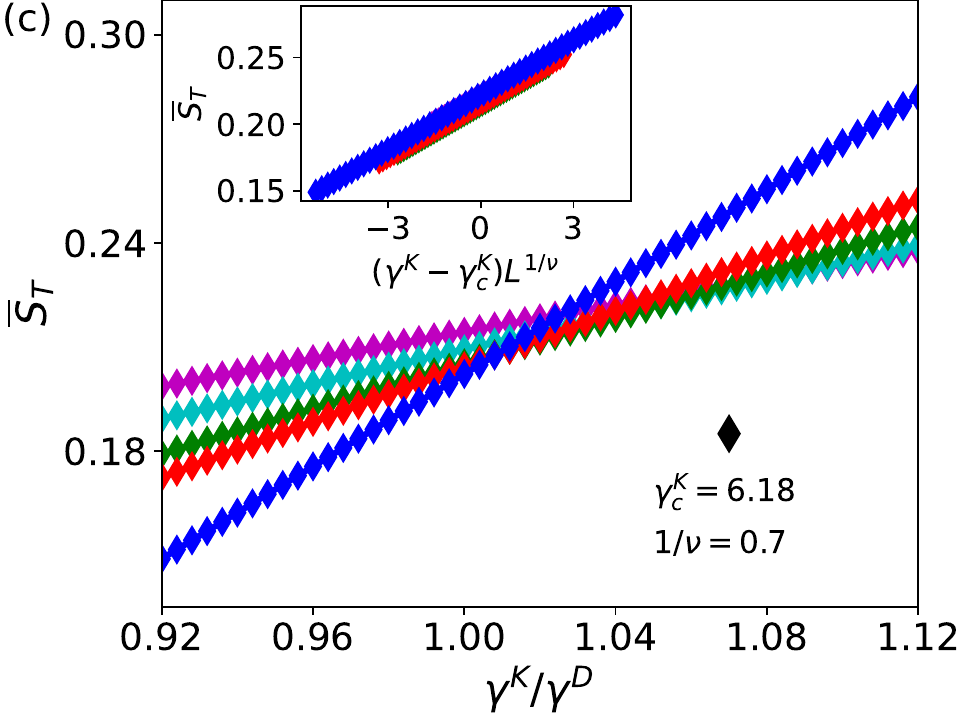}
\includegraphics[scale=0.3575]{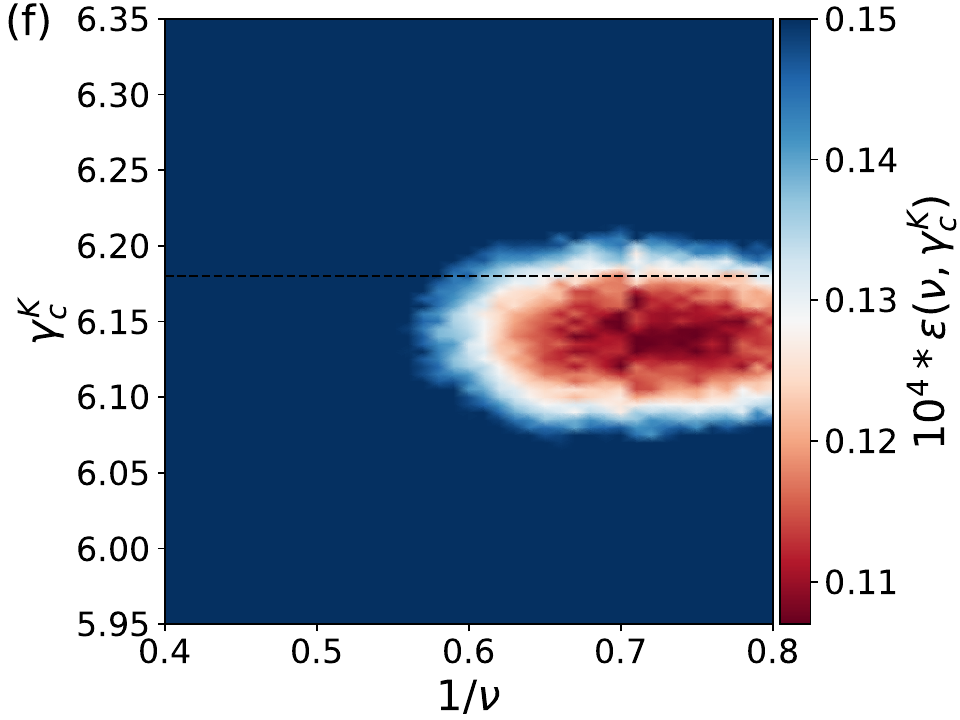}
\includegraphics[scale=0.3545]{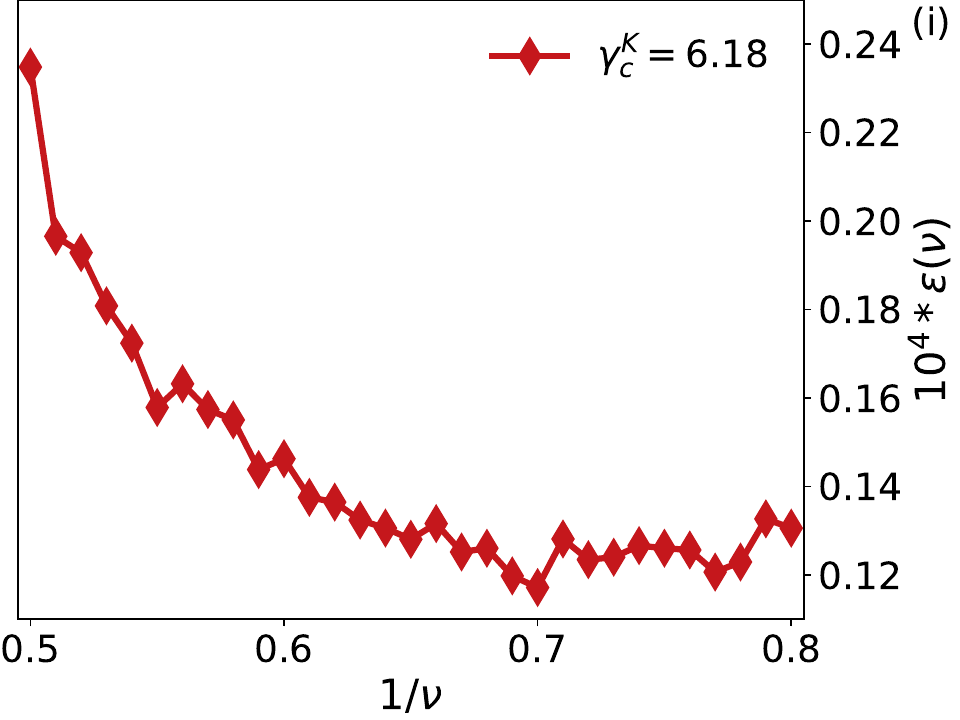}
\caption{(a,b,c) The averaged topological entanglement entropy ($\bar{S}_T$) in units of $\log 2$ as a function of relative measurement strength ($\gamma^K/\gamma^D$) and fixed rate ($p^K=p^D=0.5p^S=1, 0.833, 0.667$) of the two measurements $\hat{M}^K$ and $\hat{M}^D$ for different system sizes. (d,e,f) The objective function ($\epsilon(\nu)$) derived from the fitted data around the critical point $\gamma^K_c$ is depicted in Fig.~\ref{fig_result}(a,c,e) in the main article. The corresponding objective function for the dotted line in (d,e,f) is illustrated in (g,h,i), respectively.  All the inset shows the scaling across the critical points ($\gamma^K_c$) with the corresponding scaling exponents ($\nu$). The details of different parameters are the same as Fig.~\ref{fig_result}(a,c,e) in the main article, i.e., ($\blacksquare$) $\gamma^D=1$, $p^D=p^K=1$, ($\bigstar$) $\gamma^D=\gamma^K=3$, $p^K+p^D=1.667$, ($\Diamondblack$) $\gamma^D=\gamma^K=6$, $p^K+p^D=1.333$. }
\label{fig_fit}
\end{figure}

 Fig.~\ref{fig_fit} (a,b,c) shows a linear fit to the data in   Fig.~\ref{fig_result}(a,c,e) of the main article, in the vicinity of the critical point, and for the same choice of parameters.  The critical point, $\gamma_c^K $ is obtained by taking the average crossing point of different system sizes of the unfitted data.
Fig.~\ref{fig_fit} (d,e,f) shows the objective function calculated from the fitted data in Fig.~\ref{fig_fit} (a,b,c)  as a function of critical point $\gamma_c^K $ and scaling exponent $1/\nu $. 

It is observed that the scaling exponents ($\nu$) and critical points ($\gamma^K_c$) which minimize the objective function of the fitted data, follow a similar behavior as in Fig.~\ref{fig_intro}(a) in the main article, namely, the correlation length exponent $\nu$ decreases monotonically with the increase in measurement strengths.
Fig.~\ref{fig_fit} (g,h,i) shows a line cut of the objective function taken at the critical point corresponding to the average crossing point in Fig.~\ref{fig_fit} (a,b,c).

\newpage
\bibliography{arxiv_final}
\end{document}